\documentclass[12pt,titlepage]{article}
\pdfoutput=1
\usepackage[authoryear,round]{natbib}
\usepackage{amsmath,amssymb,graphicx,cancel,csquotes,multirow}
\usepackage[noend]{algorithmic}
\graphicspath{{graphics/}}

\usepackage[unicode,backref=page,hyperfigures=true,bookmarksnumbered=true]{hyperref}

\hypersetup{  
  pdftitle={Adjusting for Network Size and Composition Effects in Exponential-Family Random Graph Models},  
  pdfauthor={Pavel N. Krivitsky, Mark S. Handcock, and Martina Morris},  
  pdfkeywords={network size; ERGM, random graph; egocentrically-sampled data}  
} 

\DeclareMathOperator{\logit}{logit}
\DeclareMathOperator{\ilogit}{logit^{-1}}
\DeclareMathOperator{\Poisson}{Poisson}
\DeclareMathOperator{\Bernoulli}{Bernoulli}
\DeclareMathOperator{\E}{E}
\DeclareMathOperator{\Uniform}{Uniform}
\def\dysY{\mathbb{Y}}
\def\netsY{\mathcal{Y}}
\def\attrYi{\attrY_i}
\def\attrYj{\attrY_j}
\def\attrYij{\attrY\sij}
\def\attrY{x}
\def\AttrY{X}
\def\Yij{Y\sij}
\def\yij{y\sij}
\def\sij{_{i,j}}
\def\half{\frac{1}{2}}
\DeclareMathOperator{\Prob}{Pr}
\def\Peg{\Prob_{\eta,g}}
\def\ceg{c_{\eta,g}}
\DeclareMathOperator{\Odds}{Odds}
\def\pij{(i,j)}
\newcommand{\ofsize}[1]{^{(#1)}}
\def\ofsizen{\ofsize{n}}
\newcommand{\innerprod}[2]{{#1}\cdot{#2}}
\newcommand{\proglang}[1]{\textsf{#1}}
\def\ErRe{Erd\H{o}s-R\'{e}nyi}

\providecommand{\abs}[1]{\left\lvert#1\right\rvert}

\newcommand{\pkg}[1]{\texttt{#1}}
\newcommand{\myexp}[1]{\exp\left(#1\right)}
\newcommand{\ead}[1]{\texttt{#1}}

\title{Adjusting for Network Size and Composition Effects in Exponential-Family Random Graph Models}

\author{Pavel N. Krivitsky\\
Department of Statistics and iLab \\
Carnegie Mellon University, Pittsburgh, USA
\\Institute for Systems and Robotics,\\
Instituto Superior T\'{e}cnico, Lisbon, Portugal\\
\ead{pavel@stat.cmu.edu}
\and
Mark S. Handcock\\
Department of Statistics, \\
University of California at Los Angeles, Los Angeles, USA\\
\ead{handcock@stat.ucla.edu}
\and
Martina Morris\\
Department of Sociology and Department of Statistics, \\
University of Washington, Seattle, USA\\
\ead{morrism@u.washington.edu}
}

\begin{document}

\maketitle

\begin{abstract}
  Exponential-family random graph models (ERGMs) provide a principled
  way to model and simulate features common in human social networks,
  such as propensities for homophily and friend-of-a-friend triad
  closure. We show that, without adjustment, ERGMs preserve density as
  network size increases. Density invariance is often not appropriate
  for social networks. We suggest a simple modification based on an
  offset which instead preserves the mean degree and accommodates
  changes in network composition asymptotically. We demonstrate that
  this approach allows ERGMs to be applied to the important situation
  of egocentrically sampled data. We analyze data from the National
  Health and Social Life Survey (NHSLS).

\vspace*{.3in}
\end{abstract}

\section{Introduction}

Networks are a device to represent relational processes and data, that
is, data that include both the attributes of the individual units
(nodes) and the attributes of the relations (links) between
them. Examples of relational processes include the behavior of
epidemics, the interconnectedness of corporate boards, genetic
regulatory interactions, and computer networks. In social networks,
each node represents a person or social group, and each tie or edge
represents the presence or absence, or strength of a relationship
between the nodes.  Nodes can be used to represent larger social units
(groups, families, organizations), objects (airports, servers,
locations), or abstract entities (concepts, texts, tasks, random
variables).

In this paper we consider stochastic models for networks, and
Exponential-family Random Graph models (ERGMs) in particular.  This
class of models allows complex social structure to be represented in
an interpretable and parsimonious manner \citep{holland1981efp,
  frank1986mg}.  The model is a statistical exponential family for
which the sufficient statistics are a set of functions of the
network. The statistics are chosen to capture the way in which the
structure of the network departs from a simple random graph in which
the state of relationship between each pair of actors is independent
from that of every other and has a probability of $1/2$ of there being
a tie
\citep{holland1981efp,wasserman1996lml,hunter2006ice}.  Examples of
such features include long-tailed degree distributions
\citep{hamilton2008dds}, homophily, where actors prefer to associate
with actors like themselves \citep{mcpherson2001bfh,koehly2004efm},
triad-closure bias \citep{frank1986mg,snijders2006nse}, and more
complex features \citep[for example]{robins2009ccd}. 

One of the disadvantages of sophisticated models for complex
networks is that the sample space is a set of \emph{whole networks},
rather than actors in the network or dyads. This means that the
model fit based on one observed network from a population
typically cannot be directly used to infer to a population
based on a different set of actors, 
particularly if those actors differ from those
in the original network in ways which are relevant to the model.

In particular, this means that given two social networks with
different numbers of actors or different distributions of any
exogenous actor attributes relevant to the model it may not be
possible to fit the model to both of them and directly compare the
estimated parameters. Conversely, having fit an model to a particular
network, attempting to apply the model and estimated parameters to
simulate a network over a different set of actors may lead to network
structure bearing little semblance to what would be realistically
expected. For example, the natural (canonical) parametrization of an
ERGM preserves the density of a network (the ratio of the number of
ties to the number of possible dyads) as the size of the network
increases.  This implies that the number of ties per actor increases
proportionally, without limit. \citet*{anderson1999isd} studied a
similar problem with graph-level indices such as degree
centralization, by simulating the distributions of these indices on
\ErRe{} graphs of varying sizes and
densities. \citet*{goodreau2008bff} fit several exponential random
graph models to friendship networks of 59 schools of the Add Health
survey \citep{udry2003nls}. The schools varied in size from 71 to
2,209 students, and the authors briefly considered the relationship
between school size and the resulting parameter estimates. We compare
our results with those of \citet{goodreau2008bff} in
Section~\ref{sec:discussion}.

Of course, what would be considered \enquote{realistic} depends on the
specific domain in which the network is observed: some networks show
continual increase in average degree of an actor as they become larger
\citep{leskovec2007ged}, while others imply a fairly constant value
\citep{morris1991lmf,koehly2004efm}, other things being equal. In this
paper, we focus on networks of people representing personal
relationships --- friendships, sexual partnerships, gift-giving, etc.,
and our discussion will apply primarily to those. Using ERGMs to
analyze these types of data often presents a separate but related
challenge: whereas ERGMs generate probability distributions for the
dyad census --- the state of every potential relationship, such data
are extremely difficult to collect in large, sparse networks where
collection cannot be automated (such as sexual partnership networks) and
often come with severe confidentiality issues: the authors are aware
of only two sexual network datasets aiming at a dyad census: the
Colorado Springs Study \citep{woodhouse1994msn,klovdahl1994sni}, in
which a dyad census was observed among the 595 individuals ultimately
interviewed; and a census of residents of Likoma Island, Malawi, aged
18--35, interviewing them about their sexual partnerships and matching
those up to the list of island
residents \citep*{helleringer2007sns}. These are
the exceptions that prove the rule, however: in the Colorado Springs
Study, the respondents nominated a total of 5162 contacts, so most of
the individuals in this sexual partnership network were not
interviewed, while the Likoma Island study's circumstances are fairly
unique.

This is related to the network size problem in that egocentric data
typically comprise a sample of actors from the network of interest.
These data clearly contain information about some aspects of the
structure of this network, but because they are only a subset of its
actors, to infer its structural properties requires a theory on how
they are affected by network size.

We start to address these issues by discussing, in
Section~\ref{sec:wanted}, the desirable properties for a model for
social networks that would take into account network size and
composition. In Section~\ref{sec:ergm}, we show which of these properties
ERGMs do and do not have.  In Section~\ref{sec:offset}, we propose an
offset term to adjust for network size, and show how the resulting
model possesses the properties we desire.

In Section~\ref{sec:egodata}, we develop an approach to fitting ERGMs
to egocentrically sampled data from network processes that fulfill the
heuristics described in Section~\ref{sec:wanted}, by constructing
networks of varying sizes but similar structure from these
data. Finally, in Section~\ref{sec:example} we test our approach by
fitting models to constructed networks of varying sizes but similar
structure to confirm that the parameter estimates are comparable.

\subsection{Notation}
In this paper, we restrict our attention to networks of binary
relations. Thus, a network may be considered to be a set of ties. 

So for a network with $n$ actors, labeled $1, 2, \dotsc, n$, define
$\dysY\ofsizen\subseteq \{1, \dotsc, n\}^2$ to be the set of all dyads
(i.e. maximal set of ties) if relation of interest is directed
(e.g. friendship), with pairs $\dysY\ofsizen\subseteq \{\{i,j\}:
\pij\in\{1, \dotsc, n\}^2\}$ being unordered if the relation of
interest is undirected (e.g. sexual partnership). We further restrict
attention to spaces of networks where there are no constraints on the
set of potential relations of interest beyond a prohibition on
self-loops, and that have no structural constraints beyond the
constraint on the set of dyads in the network. That is,
$\netsY\ofsizen$, the set of possible networks of interest, equals to
$2^{\dysY\ofsizen}$, the power set of the possible ties. These
restrictions exclude bipartite networks, though, as we show in
Section~\ref{sec:example} networks with within-group density much
lower than between-group density can still be accommodated. Also,
while our focus is on networks with undirected ties, all of our
reasoning applies equally to networks with directed ties. We will drop
\enquote{$(n)$} where only a space of networks of a single size is
considered.

Let $\attrY$ be exogenous information --- those attributes that actors
in the network might have that may influence the structure of the
network (referred to as $\attrYi$). For the purposes of this paper, we
assume $\attrY$ to be fixed, and for brevity, we assume that any
relevant \emph{dyadic attributes} $\attrYij$ can be derived from
$\attrYi$ and $\attrYj$ and do not need to be enumerated explicitly.

For a realization $y\in\netsY$, let $\yij=1$ if the relation
of interest is present between actor $i$ and actor $j$ and $0$
otherwise, and $y_i$ be the set of neighbors of $i$ --- those actors
to which $i$ has ties. Define $y+\pij $ to be the network $y$ with a
tie between $i$ and $j$ added (if absent) and $y-\pij $ to be the
network $y$ with a tie between $i$ and $j$ removed (if present).

Throughout this paper, it is often necessary to specify precisely on
which elements of a network --- which dyads' values and attributes of
which actors --- a particular network statistic may depend. We use a
variant of the set-builder notation to do this: for example,
\enquote{$\attrYj:j\in y_i$} refers to the attributes of the neighbors
of actor $i$ and \enquote{$y_{u,v}:(u,v)\in y_i\times y_j$} refers to
the states of those dyads one of whose incident actors has a tie to
$i$ and the other to $j$.

\section{\label{sec:wanted}Desirable properties of invariant models}
In this section, we discuss what properties a network model that takes
into account network size and composition should have. In other words,
what probability model $\Prob(Y=y|\attrY;\theta)$ (that is,
probability over $y\in\netsY$ for a particular network size and
composition represented by $\attrY$, parametrized by $\theta$) would
result in similarly structured networks for similar values of
$\theta$, across different values of $\attrY$?

Answering this question empirically for social networks is fraught
with circular logic: examining what makes two networks that differ in
size and composition have similar structure requires postulating that
two or more networks over different sets of actors have similar
structure, which, in turn, requires one to postulate what similarly
structured networks look like. Thus, we focus on the local properties
of networks, and describe several heuristics that should let us
evaluate models.
\subsection{Locality}
Social processes that produce networks of human social relationships
are primarily local in nature: ties are formed and dissolved based on
the network from the point of view of the actors involved. For
example, an actor may be motivated to seek another partner by the
actor's own lack of partners, but not by a low average
number of partners in the network.

The model for a network of such actors should behave similarly: any
global network structure should be a product of local behavior and
constraints. Conversely, if the network structure, from the point of
view of an individual actor, does not change, neither should the
actor's local behavior. \citep{pattison2002nms,snijders2006nse}

\subsection{Degree distribution under scaling without composition changes}
Because each human actor typically has a finite amount of resources to
devote to relations of interest, other things being equal, adding more
actors to the network past a certain point should not substantially
increase the degree. Thus, we choose to focus on models which produce
declining marginal impacts of network size on degree distributions.

\subsection{Mixing properties}
Often, networks of interest are not homogeneous --- actors may possess
attributes, such as sex, socioeconomic status, and age that influence
with whom they associate. Counts of ties broken down by attributes of
the actors involved --- called mixing matrices --- have been modeled
using log-linear models, where they have been presented as a function
of the numbers of actors with each attribute value, overall
attribute-specific propensities of actors with each attribute value to
form ties, and an additional \enquote{selectivity} factor representing
the propensity of actors from each attribute class to form ties with
each other. \citep{morris1991lmf}

From the point of view of an individual actor, this suggests that an
actor's degree should be a function of how that actor's own affinities
match up with the attribute composition of the population, which
affects how often an actor might encounter potential partners with the
actor's preferred attributes.

\section{\label{sec:ergm}Structure of exponential-family random graph models}
We now discuss how well ERGMs fulfill these criteria. Consider a
general curved ERGM for networks of binary relations
\citep{hunter2006ice},
\begin{equation}
\Peg(Y=y|\attrY;\theta)=\frac{\myexp{\innerprod{\eta(\theta,\attrY)}{g(y,\attrY)}}}{\ceg(\theta,\attrY)},\ y\in\netsY, \label{eq:cergm}\end{equation} with 
\[\ceg(\theta,\attrY)=\sum_{y'\in\netsY} \myexp{\innerprod{\eta(\theta,\attrY)}{g(y',\attrY)}},\]
where $g(\cdot,\cdot)$ is a vector of sufficient statistics (also
incorporating exogenous information $\netsY$), $\theta$ is a vector of
model parameters, $\eta(\cdot,\cdot)$ is a mapping from the model
parameters $\theta$ (also incorporating exogenous information
$\attrY$) to natural parameters , and $c_{\cdot,\cdot}(\cdot,\cdot)$
is the normalizing constant. (Often, $\eta(\theta,\attrY)=\theta$ for
a \emph{linear ERGM}.)  Depending on $g$, it may be
intractable. \citep{hunter2006ice}

\subsection{\label{sec:change-statistics}Change statistics}
An interpretation of an ERGM from the point of view of individual
actors comes in the form of \emph{change statistics}. A change
statistic of a network statistic $g_k$ is the change in its value
associated with toggling a dyad (say, $\pij$),
\[\Delta\sij g_k(y,\attrY)\equiv g_k(y+\pij ,\attrY)-g_k(y-\pij ,\attrY).\]
The conditional probability of a tie between $i$ and $j$ given the
rest of the network is a function of the change statistics for $\pij$,
reduces, through cancellations, to
\[\Peg(\Yij=1|\attrY,Y-\pij =y-\pij ;\theta)=\ilogit\left(\innerprod{\eta(\theta,\attrY)}{\Delta\sij g(y,\attrY)}\right),\]
for $\ilogit(x)\equiv \frac{1}{1+\myexp{-x}}$. (See
Appendix~\ref{app:change-stat-deriv} for the complete derivation.)

Consider a hypothetical discrete Markov process in which, during each
step, a pair of actors $\pij\in\dysY$ is selected at random, and they
either form (or maintain) a tie between them, with probability
$\ilogit\left(\innerprod{\eta(\theta,\attrY)}{\Delta\sij
  g(y,\attrY)}\right)$ or dissolve (or maintain absence of) a tie
between them otherwise.

A selection of $i$ and $j$ may be viewed as an \enquote{opportunity}
for actors $i$ and $j$ to have or not to have a tie and the
\enquote{decisions} by these actors whether or not to do so may be
viewed as made based on the factors that the actors take into
consideration ($\Delta\sij g_k(y,\attrY)$) and how they weigh these
factors ($\eta_k(\theta,\attrY)$). Thus, if a change statistic does
not depend on some datum, it may be viewed as having the actors make
the decision while being ignorant of that datum or choosing not to
take that datum into account. This process is a Gibbs sampling
algorithm (formally described in Appendix~\ref{app:Gibbs-alg}) that generates a
draw from an ERGM over the space of graphs $\netsY$ having $\theta$ as
its parameters and $g$ as its sufficient statistics, so an ERG may be
viewed as a consequence of a long series of these opportunities and
decisions. \citet*{robins2001rgm} draw a similar parallel --- the
global network structures arising as a consequence of local processes.

We do not assert that this is a realistic description of a temporal
network process (if only because only one tie may be formed or
dissolved during each time step), but it serves as a useful analogy.

\subsection{\label{sec:CS-locality}Locality}
The notion of locality of a model can thus be expressed through the
dependencies of the change statistics. It is not sufficient to specify
the dependence of dyads on states of other dyads, however: it is also
important to consider dependence on attributes of the network, the
actors, and the dyads that would, under most formulations, be
considered exogenous, and thus effectively conditioned on. For
example, while a count of ties $g(y,\attrY)=|y|$, resulting in
$\Delta\sij g(y,\attrY)=1$ would be a \enquote{local} statistic,
network density $g(y,\attrY)=\frac{2|y|}{n(n-1)}$ (for undirected
networks) would not be local in this sense, because its change
statistic, $\frac{2}{n(n-1)}$, depends on the network size --- a
network-wide attribute. Using the analogy described in
Section~\ref{sec:change-statistics}, a model with a density sufficient
statistic would be akin to actors making their \enquote{decision}
based on the total number of actors in the network --- a very
non-local decision rule. The relationship between notions of social
neighborhoods and ERGM dependence structure was also discussed by
\citet*{pattison1999lml}.

We thus discuss three notions of locality, based on three types of
dyadic dependence that have appeared in literature.

\subsubsection{\label{sec:local-dyind}Locality based on dyadic independence}
If a change statistic for $\pij$ only depends on $\yij$ (and, for
directed networks, $y_{j,i}$), then the model has dyadic independence
--- all dyads are stochastically independent, and the probability of
the network is a product of individual dyad probabilities. With
respect to exogenous attributes, the corresponding constraint is that
the change statistic must also not depend on attributes of actors
other than $i$ and $j$: $\Delta\sij
g(y,\attrY)=f(\attrYi,\attrYj)$ for some function
$f(\cdot,\cdot)$. (Note that $f$ is not a function of $\yij$
itself, since it is a difference between the network $y+\pij $ and
network $y-\pij$, no matter the present state of $\yij$.) However,
the class of models with dyadic independence is fairly limited
\citep{holland1981efp,frank1986mg,snijders2006nse}, so weaker notions
of locality are needed in order for the concept to be useful.

\subsubsection{\label{sec:local-markov}Locality based on Markov dependence}
Described by \citet*{frank1986mg}, a \emph{Markov graph} is one in
which two dyads are conditionally independent given the rest of the
network unless they share an actor. In terms of change statistics, it
means that a change statistic for a dyad $\pij$ may only depend on
dyads incident on actor $i$ and dyads incident on actor
$j$. Sufficient statistics in this class that are meaningfully local
but do not preserve dyadic independence include the count of actors in
the network that have exactly $d$ ties: $g(y,\attrY)=\sum_{i=1}^n
1_{|y_i|=d}$, for an undirected network, has change statistic
\[\Delta\sij g(y,\attrY) = (1_{|(y+\pij)_i|=d} + 1_{|(y+\pij)_j|=d})-(1_{|(y-\pij)_i|=d} + 1_{|(y-\pij)_j|=d}),\]
which only depends on the dyads incident on the actors incident on the
dyad of interest.

While models described by \citet*{frank1986mg} do not make explicit
use of exogenous attributes of dyads and actors, it is often desirable
to incorporate these into the model as in the Markov block models of
\citet*{strauss1990pes}. However, directly extending the concept of Markov
dependence to exogenous dyadic and actor attributes results in a
definition that allows a dyad $\pij$ to depend on attributes of any
and all dyads $(i,k)$ and $(j,k)$, for all $k\notin\{i,j\}$ and thus
on attributes of any actor $k$. This definition is not meaningfully
local --- at least not in human networks being considered. Thus, a
useful definition of change statistic locality beyond dyadic
independence must be \emph{realization-dependent} --- that is, it must depend
on the specific configuration of the social neighborhood of the dyad
of interest.

We define a \emph{Markov graph local change statistic} as one that only
depends on states of dyads incident on $\pij$ and only on exogenous
attributes of those actors that \emph{have ties} to either
$i$ or $j$. That is,
\[\Delta\sij g(y,\attrY)=f(y_{i},y_{j},\attrYi,\attrYj,\attrY_{k}:k\in y_i\cup y_j)\]
for some function $f(\dotsb)$. The conditional dependence structure of
of the Markov graphs is thus \emph{realization-independent} with
respect to dyad values in that, for a given dyad $\pij$, the set of
dyads whose states may affect the conditional probability of $\pij$
having a tie does not depend on what other ties are present in the
network. On the other hand, it is \emph{realization-dependent} with respect
to the actor attributes, in the sense that $\pij$ does not depend on
$\attrY_k$, unless there is a tie between $i$ and $k$ or between $j$
and $k$.

\subsubsection{Locality based on partial conditional independence}
\citet{pattison2002nms} and \citet*{snijders2006nse} define an even
broader class of network models that still preserve the local nature
of the sufficient statistics --- \emph{partial conditional
  dependence}, a realization-dependent dependence structure for dyads,
where dyads $\pij$ and $(u,v)$ are conditionally independent given
the rest of the network unless they either are incident on the same
actor (i.e. $i=u$, $i=v$, $j=u$, or $j=v$), or if there exist edges at
both $(i,u)$ and $(j,v)$ (i.e. $y_{i,u}=y_{j,v}=1$), or vice versa.

Because dependence of change statistics directly reflects conditional
dyad dependence, this means that a change statistics for dyad $\pij$
may only be a function of the states of those dyads $(u,v)$ that
fulfill the criteria above, and a natural constraint on the exogenous
attributes on which the statistic may depend is that it may depend
only on the attributes of actors that would be involved in
the social neighborhood defined by the conditional independence: dyads
and actors that have ties to either $i$ or $j$ and dyads both of whose
incident actors have ties to $i$ or $j$. Concretely,
\[\Delta\sij g(y,\attrY)=f(y_i,y_j, y_{u,v}: (u,v)\in y_i\times y_j, \attrYi,\attrYj,\attrY_{k}:k\in y_i\cup y_j)\]
for some function $f(\dotsb)$.

Thus, by choosing appropriate change statistics, an ERGM can be made
\enquote{local}, and this class of statistics is fairly rich, including
$k$-star, degree, and triangle counts \citep{frank1986mg}, mixing
terms \citep{koehly2004efm}, and shared partner
distributions \citep{snijders2006nse,hunter2006ice}.

\subsection{\label{sec:scale-ergm}Scaling without composition changes}
However, any linear ERGM suffers from a problem: if
$\eta(\theta)=\theta$, then it can be shown that for any
$g(\cdot,\cdot)$, if $\theta$ is set to give an average degree of
$\mu$ for a particular number of actors $n$, then for a different $n$,
the expected average degree will be different from $\mu$ under this
model.

Intuitively, this is because for a network to maintain the same mean
degree, the number of ties must grow linearly in the number of
actors. However, for a constant value of $\theta=0$, the network
distribution always reduces to an \ErRe{} graph with density $\half$,
whose expected number of dyads grows quadratically in the number of
actors, so the mean degree inevitably increases. A more rigorous
derivation of this is given in the Appendix~\ref{app:scale-ergm-proof}.

In the following section, we consider adjusting the ERGM for network
size effects.

\section{\label{sec:offset}An offset model to adjust for network size}
In this section, we consider adding a single offset term to the ERGM
to adjust the model for network size effect.  An \emph{offset} term is
a component of the vector $g(\cdot,\cdot)$ which does not have a free
parameter associated with it. The coefficient of the term is instead a
known constant or a function of known quantities. This terminology is
extended from that for Generalized Linear Models
\citep[p. 206]{mccullagh1989glm}.

\subsection{Model statement}
Specifically, we add a term that would ensure that mean degree would converge, asymptotically, in the absence
of all other terms:
\[\Peg(Y\ofsizen=y|\attrY\ofsizen;\theta)=\frac{\myexp{\log\left(\frac{1}{n}\right) \abs{ y}+\innerprod{\eta(\theta,\attrY\ofsizen)}{g(y,\attrY\ofsizen)}}}{\ceg(\theta,\attrY\ofsizen)},\ y\in\netsY\ofsizen\]
\[\ceg(\theta,\attrY\ofsizen)=\sum_{y'\in\netsY\ofsizen} \myexp{\log\left(\frac{1}{n}\right)\abs{ y'}+\innerprod{\eta(\theta,\attrY\ofsizen)}{g(y',\netsY\ofsizen)}}.\]
Here, $n$ is, again, the number of actors in the network.

There is an intuitive interpretation for the offset term,
suggested by the Na\"{i}ve Gibbs sampling and change statistics from
Section~\ref{sec:change-statistics}. Recall that the conditional
log-odds of a tie at $\pij$ given the rest of the network is
\begin{multline*}
\logit \left(\Peg(\Yij\ofsizen=1|\attrY\ofsizen,Y\ofsizen-\pij =y-\pij ;\theta)\right)=\\\innerprod{\eta(\theta,\attrY\ofsizen)}{\Delta\sij g(y,\attrY\ofsizen)}.\end{multline*}
In the presence of the offset term, this becomes
\begin{multline*}
\logit \left(\Peg(\Yij\ofsizen=1|\attrY\ofsizen,Y\ofsizen-\pij =y-\pij ;\theta)\right)=\\\log \frac{1}{n} + \innerprod{\eta(\theta,\attrY\ofsizen)}{\Delta\sij g(y,\attrY\ofsizen)},
\end{multline*}
or
\begin{multline*}
\Odds_{g,\eta}(\Yij\ofsizen=1|\attrY\ofsizen,Y\ofsizen-\pij =y-\pij ;\theta)=\\\frac{1}{n} \myexp{ \innerprod{\eta(\theta,\attrY\ofsizen)}{\Delta\sij g(y,\attrY\ofsizen)}},
\end{multline*}
so the conditional odds of each dyad given the rest of the network are
multiplied by $\frac{1}{n}$. This can be viewed as reflecting the
declining fraction of the network with which each actor may get an
\enquote{opportunity} to make contact (although this interpretation is not necessary).

Given this \enquote{opportunity}, the effect of each term
$\eta_k(\theta,\attrY\ofsizen)\Delta\sij g_k(y,\attrY\ofsizen)$ on the
conditional log-odds of the tie does not depend on network size and
composition, since $g_k(\cdot,\cdot)$ is local.

We now describe some asymptotic properties of this model for the cases of dyadic
independence. A model with dyadic dependence is demonstrated in
Section~\ref{sec:example}.

\subsection{\label{sec:offset-edge-count}\ErRe{} model}

A model with the offset term and a single edge-count term,
\begin{align}
  \Peg(Y\ofsizen=y|\attrY\ofsizen;\theta)&\propto \myexp{\log\left(\frac{1}{n}\right)\abs{ y}+\theta\abs{ y}}\label{eq:erre-offset}\\
  &\propto \myexp{\left(-\log n+\theta\right)\abs{ y}}\notag
\end{align}
results in a \ErRe{} network \citep{holland1981efp} with the probability of each individual tie, independently,
\[\Peg(\Yij\ofsizen=1|\attrY;\theta)=\ilogit\left(-\log n + \theta\right).\]
As network size $n$ increases, the probability that a given vertex has a particular degree $d$.
\[\lim_{n\to \infty} \Peg(|Y\ofsizen_{i}|=d|\attrY)=\frac{1}{d!}\left(\myexp{\theta}\right)^{d}\myexp{-\myexp{\theta}}.\]
(Full derivation is given in Appendix~\ref{app:erre-asymp}.) Thus, the
degree distribution converges to $\Poisson(\myexp{\theta})$. With the
expected mean degree converging to $\myexp\theta$, $\theta$, which, in
the absence of the offset, would determine the density of the network,
instead determines the network's mean degree. Conversely, for
sufficiently large networks with similar mean degree, the maximum
likelihood estimates of $\theta$ would be similar.

\subsection{\label{sec:mixmat}Selective mixing model}
In many circumstances, actors can be partitioned into $K$ exogenous
groups, and propensities of actors in one group to form ties to
another (or others in the same group) can be modeled using ERGMs
\citep{koehly2004efm}. We describe here how these models behave under
changing network size and how they interact with composition in the
presence of the offset. Suppose that for a sequence of random networks
$Y\ofsize{2},Y\ofsize{3},\dotsc$ of increasing size, their actor attributes
$\attrY\ofsizen$ partition the actors into a partitioning $P\ofsizen_k$,
$k=1,\dotsc,K$, with $P\ofsizen(i)$ giving $k$ such that $i \in
P\ofsizen_k$, and $y_{P\ofsizen_{k_1},P\ofsizen_{k_2}}$ being the set of
ties between actors in $P\ofsizen_{k_1}$ and actors in
$P\ofsizen_{k_2}$. Suppose that the $\attrY\ofsizen$ are such that these
proportions converge: $\lim_{n\to\infty}|P\ofsizen_k|/n=p_k$.

Consider the following mixing model \citep{koehly2004efm} for a given
size, with the proposed offset added:
\[\Peg(Y\ofsizen=y|\attrY\ofsizen;\theta)\propto \myexp{\log\left(\frac{1}{n}\right) |y|+\sum_{k_1,k_2}\eta_{k_1,k_2}(\theta)\abs{ y_{P\ofsizen_{k_1},P\ofsizen_{k_2}}}}.\]
This model is dyad-independent and local, with all dyad values for
each combination of $k_1$ and $k_2$ being identically distributed.
$\eta_{k_1,k_2}(\theta)$ can be thought of as representing preferences
of actors in group $k_1$ toward actors in group $k_2$. Different forms
of $\eta(\cdot)$ may be used to model different patterns of mixing,
such as assortative (homophily), disassortative, and even overall
group activity levels. Then,
\[\Peg(\Yij\ofsizen=1|\attrY\ofsizen;\theta)=\ilogit\left(-\log n +\eta_{P\ofsizen(i),P\ofsizen(j)}(\theta)\right)\]
and the expected degree of some actor $i$ is
\begin{align*}
\sum_{j=1}^n\Peg(\Yij\ofsizen=1|\attrY\ofsizen;\theta)&=\sum_{j=1}^n\ilogit\left(\log\frac{1}{n} +\eta_{P\ofsizen(i),P\ofsizen(j)}(\theta)\right)\\
&=\sum_{k_2=1}^K|P\ofsizen_{k_2}|\ilogit\left(-\log n +\eta_{P\ofsizen(i),k_2}(\theta)\right),
\end{align*}
which, as network size increases, becomes
\begin{align*}
\lim_{n\to\infty}\sum_{j=1}^n\Peg(\Yij\ofsizen=1|\attrY\ofsizen;\theta)&=\sum_{k_2=1}^K\left(\lim_{n\to\infty}\frac{|P\ofsizen_{k_2}|}{n}\right) \myexp{\eta_{P(i),k_2}(\theta)}\\
&=\sum_{k=1}^K p_{k} \myexp{\theta_{P(i),k}}.
\end{align*}
Thus, asymptotically, the number of ties actor $i$ in group $P(i)$ is
expected to have to actors in group $k$ (i.e. the actor's mean degree
with respect to that group) is proportional to the fraction of the
actors in the network made up by members of $k$ (i.e. how often $i$
gets an opportunity to make a tie with a $k$, relative to others) and
proportional to $\myexp{\eta_{P(i),k}(\theta)}$ (i.e. how much actor $i$
favors/disfavors ties with members of $k$). The expected overall
degree of that actor is thus a function of how well availability
($p_k$) matches up with affinities ($\eta_{P(i),k}(\theta)$).

Conversely, if $\eta(\cdot)$ is a linear transformation, the MLE for
$\theta$ would be similar for networks of different size and
composition if they had this proportional mixing structure.

\section{\label{sec:egodata}Inference from egocentrically sampled data}
An ERGM applies to a dyad census --- the enumeration of dyad states of
all dyads among a particular set of actors. Egocentrically sampled
data --- data collected by surveying a sample of the actors
(\enquote{egos}) about actors to whom they are tied in the network of
interest (\enquote{alters}) \citep{koehly2004efm} --- only contains
information about dyads incident on each of the sampled
egos. Furthermore, alter identities are not observed, only their
attributes of interest are, so it is not known, for example, whether
two egos reporting two alters with similar attributes are, in fact,
referring to the same individual, and whether two egos who each
describe an alter with attributes similar to those of the other ego
are, in fact, referring to each other. In order to analyze
egocentrically sampled data using ERGMs, we consider hypothetical full
networks from which the egocentrically observed egos could have come,
and what their network statistics of interest would have to have been
in order to have produced the egocentric data that were observed.

\subsection{\label{sec:ego-meanstats}Deriving sufficient statistics from an egocentric census}
We describe how certain network statistics can be computed from
a census of the egos in an observed network.
Because they can only depend on
information about actors in the sample and their immediate neighbors,
they are local according to the Markov graph variant of the
definition, given in Section~\ref{sec:local-markov}.

Let $E$ be the set of egos (respondents) and $A_e$ be the set of
alters (nominations) nominated by ego $e\in E$. Note that these are
nominations, rather than actors: a single actor may be nominated
multiple times and egos may nominate each other, and they all appear
as distinct nominations. Lastly, let $\attrY_e$ and $\attrY_a$ be
attributes of interest of the respective egos and alters. Define
$A=\bigcup_{e\in E}A_e$.

\subsubsection{Dyad census statistics}
When an undirected network is observed egocentrically, with a census
of actors, each tie is reported twice: once by each of the actors
involved. Thus, a dataset with $\abs{ A}$ alters nominated by $\abs{
  E}$ egos could have been observed on a network of $\abs{ E}$ actors
and a total of $\frac{\abs{ A}}{2}$ ties.

More generally, a network statistic that is the summation over the
edges in the network of some function $f(\attrYi,\attrYj)$ of
the attributes of the actors incident on the edge,
\begin{equation}
g_k(y,\attrY) = \sum_{\pij\in \dysY} \yij f(\attrYi,\attrYj),\label{eq:sumstat}
\end{equation}
would be observed egocentrically as $f(\attrY_e,\attrY_a), e\in E,
a\in A_e$, with each tie being observed twice: once when $i$ had
nominated $j$ and once when $j$ had nominated $i$. Thus,
$g_k(y,\attrY)$ is
\begin{equation}
  \frac{1}{2} \sum_{e\in E} \sum_{a\in A_e} f(\attrY_e,\attrY_a).\label{eq:suminfer}
\end{equation}

Sufficient statistics for a selective mixing model such as that in
Section~\ref{sec:mixmat}, the count of ties between actors of a
particular pair of categories (say, $k_1$ and $k_2$) can be expressed
in the form \eqref{eq:sumstat}, for 
\[f(\attrYi,\attrYj)=
\begin{cases}
  1_{i\in P_{k_1}}1_{j\in P_{k_1}}& \text{if $k_1=k_2$}\\
  1_{i\in P_{k_1}}1_{j\in P_{k_2}}+1_{j\in P_{k_1}}1_{i\in P_{k_2}}& \text{otherwise}\\
\end{cases},
 \]
with the second case being a consequence of the network being
undirected. Observing the network egocentrically, ties counted in
the case of $k_1=k_2$ are reported twice as ties between $k_1$ and $k_1$,
while ties counted in the $k_1\ne k_2$ case are reported as one partnership
with the ego in $k_1$ and the alter in $k_2$ and another partnership
with the ego in $k_2$ and the alter in $k_1$. Thus, the 
number of ties between $k_1$ and $k_2$ is
\[
\begin{cases}
  \frac{1}{2}\sum_{e\in E} \sum_{a\in A_e} 1_{e\in P_{k_1}}1_{a\in P_{k_1}}& \text{if $k_1=k_2$}\\
  \frac{1}{2}\sum_{e\in E} \sum_{a\in A_e} 1_{e\in P_{k_1}}1_{a\in P_{k_2}}+1_{j\in P_{k_1}}1_{i\in P_{k_2}}& \text{otherwise}
\end{cases}.
 \]

\subsubsection{Actor census statistics}
Statistics such as the number of actors with a particular degree (or range
of degrees) or the number of $k-\text{stars}$ that are local are observed
directly in an egocentric census: they are the properties of
the egos. Thus, they are statistics that can be expressed as a summation over
the actors of some function of each actor and its neighbors:
\begin{equation}
  g_k(y,\attrY) = \sum_{i=1}^n f(\attrYi,\attrYj:j\in y_i).\label{eq:sum-actors-stat}
\end{equation}
Suppose $f(\attrYi,\attrYj:j\in y_i)$ is local --- that is, it only
depends on exogenous properties of actor $i$ and actors $j\in y_i$.
Then $f(\attrYi,\attrYj:j\in y_i)$
would be egocentrically observed as $f(\attrY_e,\attrY_a:a\in A_e),
e\in E$. Thus statistics of the form~\eqref{eq:sum-actors-stat} can be
expressed as $\sum_{e\in E} f(\attrY_e,\attrY_a:a\in A_e)$.

In particular, the count of actors with a particular degree $d$,
$g_k(y,\attrY)=\sum_{i=1}^n 1_{\abs{ y_i} = d}$ can be expressed in
the form of~\eqref{eq:sum-actors-stat} with $f(\attrY_e,\attrY_a:a\in
A_e)=1_{\abs{A_e }=d}$, so the number of actors with degree
$d$ is simply $\sum_{e\in E} 1_{\abs{ A_e} = d}$.

\subsection{\label{sec:consistency}Sampling and consistency}
Section~\ref{sec:ego-meanstats} describes the derivation of
sufficient statistics from egocentric data consisting of \emph{all}
the actors in the network of interest, rather than a sample of actors,
and the data available are a sample. However, if the sample of egos is
representative (i.e. is a simple random sample or a properly weighted
stratified sample), the distribution of egos and ego reports in a
network is representative of those in the full network at the time the
data were collected. In particular, the degree
distribution in the sample is representative of that in the full
network and the selective mixing observed in the sample is
representative of that in the full network, provided that the
underlying network process is local.

Another potential problem arising when inferring network statistics
from sampled egocentric data, as opposed to a census of all actors in
the network, is possible mutual inconsistency of reports. For example,
consider an undirected network of sexual partnerships such as the one
modeled in Section~\ref{sec:example}. Assuming no nonresponse and
truthful reports, egocentric census of such a network would produce
the same total number of ties to females reported by males as ties to
males reported by females, and the total number of partnerships
reported by all actors would necessarily be even (as each
partnership is reported twice). An egocentric sample will not
necessarily produce mutually consistent reports, and it may be the
case that no network having the exact statistics can be
constructed, either because reports are inconsistent or because a
fractional number of ties is implied.

While there is ongoing work on more sophisticated approaches to deal
with inconsistent reports \citep[for example]{admiraal2009dnm}, we
take the simple approach of taking the average of conflicting reports:
in the above example, if the number of ties to females reported by
males is different from the number of ties to males reported by
females, we use their average as the implied number of male-female
ties, as~\eqref{eq:suminfer} suggests. Also, from the point of view
of ERGM-based inference and simulation, an implied fractional number
of ties is not problematic: instead of considering the value
a statistic of a concrete network, we may consider it a mean-value
parameter, the expected value of the network statistic in question
under the distribution whose (natural) parameters are of interest
\citep{duijn2009fcm}.

The network inferred from egocentrically sampled data has a degree
distribution similar to that of the population to the extent that the
egocentric sample is representative of it, and its mixing properties
are similar as well: the composition of the inferred network is
proportional to that of the sample and thus approximately proportional
to that of the population, and the average number of relations in the
inferred network that an actor with a particular value of a given
attribute has with actors of a particular (possibly different) value
of a given (possibly different) attribute is close to that of the
sample and thus that of the population. Therefore, according to our
heuristics the structure of the inferred network is at least
approximately similar to that of the full network, which suffices for
the following demonstration.

\section{\label{sec:example}Application to National Health and Social Life Survey data}
In this section, we illustrate the approach in the context of real
data. To examine whether a model has desirable properties with respect
to networks of varying sizes and compositions requires postulating two
or more distinct networks as having the same structure, and we
construct networks of increasing size but with similar structure by
extrapolating data from the 1992 National Health and Social Life
Survey (NHSLS). \citep{laumann1994sos,laumann1992nhs}

For each of a range of network sizes, we use a bootstrap of the egos
(with their nominations) in the sample to generate a pseudo-population
of egocentric network datasets and network statistics
implying, on average, similar structure according to the criteria of
Section~\ref{sec:wanted}, and fitting the same model to each of these
networks, to see if the results from the model are comparable across
network sizes. We elaborate on the exact procedure below.

\subsection{NHSLS Data}
The data comprise a stratified random sample of 3,432 American men and
women between 18 and 60 years old. Respondents were asked to report on
all of the spousal or cohabiting partnerships they had ever had, and
all of the sexual partnerships they had had in the last year.  For the
purposes of this analysis, we focus only on the partnerships that were
active on the day of the interview.  As a result, any respondent
reporting more than one active partnership can be defined as having
``concurrent'' partnerships \citep*{morris1997cps}.  As this was an
egocentric sampling design, the partners were not enrolled in the
study.  Instead, the respondent was asked to report on many aspects of
the partner and the partnership. Among the information collected were
the age, sex (male or female), and race/ethnicity (Asian/Pacific
Islander, White, Black, Alaskan/Native American, Hispanic, or
\enquote{Other}) of each respondent and of all of the respondent's
sexual partners.

\subsection{\label{NHSLS-weighting}Missing data and weighting}

For the purpose of this analysis, the smallest racial categories ---
Asian/Pacific Islander (67 egos, 51 alters), Alaskan/Native American
(45 egos, no alters), and \enquote{Other} (no egos, 58 alters) --- were all
merged.

Five egos have missing or invalid information on age, race, or sex,
and 67 egos have missing or invalid information on age, race, or sex
of one or more of their alters. We excluded these (72) egos and their
alters from the analysis.

By design, the respondents (egos) were to be aged 18--59; however, a
few (3) of the respondents are 60, and we exclude them from the
analysis. At the same time, there was no age limit on the alters
nominated, so the youngest alter is 16 and the oldest alter is 82. The
hypothetical network we construct from these data is the network of
egos --- all 18--59 --- so to make it \enquote{closed} we exclude all
alters younger than 18 (21 alters) or older than 59 (114 alters), but
not the egos who have nominated them. Thus, we model the network of
sexual partnerships \emph{between individuals who are 18--59}.

We use the post-stratification weights provided by the study to adjust
for the design of the stratified sample and the post hoc analysis of
non-response patterns.  This ensures that both egos and alters are
proportionally represented. We give the breakdown of this population
and the weighting in Table~\ref{tab:NHSLS-summs}. All data summaries
and figures that follow incorporate these weights.
\begin{table}
  \caption{\label{tab:NHSLS-summs} Ego/actor attributes, sampling weights, and adjusted composition. (Groups with a lower sampling weight had been oversampled and/or had higher response rates than those with higher sampling weight.)}
  \centering
    \begin{tabular}{lrcrcr}
\hline
   & $\text{Respondents} $&$\times$&$ \text{ Mean weight}$ & $\propto$ & Composition \\
  \hline
  Sex &&&&& \\
  \quad Female & $1890$&&$ 0.90$ && $50.6\%$\\
  \quad Male & $1467$&&$ 1.13$ && $49.4\%$\\
  Racial category &&&&&\\
  \quad Black & $541$&&$ 0.74$ && $11.9\%$\\
  \quad Hispanic & $314$&&$ 0.98$ && $9.2\%$\\
  \quad Other & $106$&&$ 1.23$ && $3.9\%$\\
  \quad White & $2396$&&$ 1.05$ && $75.0\%$\\
  Age group &&&&&\\
  \quad $18$--$19$ & $106$&&$ 1.32$ && $4.2\%$\\
  \quad $20$--$29$& $933$&&$ 0.99$ && $27.6\%$\\
  \quad $30$--$39$& $1060$&&$ 0.91$ && $28.8\%$\\
  \quad $40$--$49$& $747$&&$ 1.09$ && $24.2\%$\\
  \quad $50$--$59$& $511$&&$ 0.99$ && $15.1\%$\\
\hline
\end{tabular}
\end{table}

In order to generate a network dataset with a particular number of
egos, we sample egos, with replacement, from the set of NHSLS
respondents 18--59, with missing data treated as above, weighted by
the sampling weights. If the reweighted NHSLS survey data are
considered to be the empirical distribution of the target population,
this approach is a form of nonparametric bootstrap
\citep{davison1997bma}.

\subsection{\label{sec:NHSLS-model}Modeling the sexual partnership network}
We model the hypothetical network that could have produced the
resampled egocentric data as a linear ERGM, with the sufficient
statistics reflecting actor attributes as we expect them to affect the
network structure.  The model includes terms that represent the
effects of three nodal attributes --- sex, race, and age --- and
propensities for monogamy.  All of the statistics are local in the
sense defined in Section~\ref{sec:CS-locality}: the monogamy
propensity statistics have Markov dependence structure
(Section~\ref{sec:local-markov}), and the others are dyad-independent
(Section~\ref{sec:local-dyind}).

\subsubsection{Sex}
Under our model, the propensity to have partners is affected by sex in
three major ways. Firstly, different sexes may have different overall
propensities to have partners, and different degree of propensity
toward monogamy (Table~\ref{tab:NHSLS-degdist-sex}). Secondly,
same-sex partnerships are rare
(Table~\ref{tab:NHSLS-mixing-sex}). Finally, in heterosexual
partnerships, the male partner is often older than the female
partner. (In this dataset, in heterosexual partnerships, the female
partner is, on average, $1.8$ years younger than the male partner.)  We
model these effects by adding the following sufficient statistics to
the ERGM:
\begin{description}
\item[overall
  propensity of actors of each sex to have ties] represented by the number of partners of actors of each sex:
  \[g_{[1,2]}(y,\attrY)=\left(\sum_{i\in \text{Female}}\abs{ y_i},\sum_{i\in \text{Male}}\abs{ y_i}\right),\]
  (and note that $\sum_{i\in \text{Male}}\abs{ y_{i}}+\sum_{i\in
    \text{Female}}\abs{ y_{i}}=2\abs{ y}$);
\item[relative prevalence of same-sex partnerships] represented by the number of same-sex ties:
  \[ g_{3}(y,\attrY)=\abs{ y_{\text{Female},\text{Female}}}+\abs{ y_{\text{Male},\text{Male}}};\]
\item[propensity toward monogamy] represented by the number of actors of each sex having exactly one partner:
  \[g_{[4,5]}(y,\attrY)=\left(\sum_{i\in \text{Female}} 1_{\abs{ y_i}=1},\sum_{i\in \text{Male}} 1_{\abs{ y_i}=1}\right);\]
\item[age-sex asymmetry in partnerships] represented by the number of ties between an older male and a younger female:
  \begin{multline*}g_{19}(y,\attrY)=\sum_{\pij\in
    \dysY}\yij\left(1_{\text{($i$ is Male, $j$ is Female, and
      $t_i>t_j$)}}+\right.\\\left.1_{\text{($j$ is Male, $i$ is Female, and
      $t_j>t_i$)}}\right),\end{multline*} where $t_i$ is age of actor $i$.
\end{description}

\begin{table}
  \caption{\label{tab:NHSLS-degdist-sex} Reported actor degree distribution, by sex}
  \centering
  \begin{tabular}{lrrrrrr}
    \hline
    Degree & $0$ & $1$ & $2$ & $3$ & $4$ & Mean\\
    \hline
    Female & $29.6\%$ & $69.0\%$ & $1.4\%$ & $0.0\%$ & $0.0\%$ &$0.72$\\
    Male & $24.3\%$ & $71.0\%$ & $4.5\%$ & $0.2\%$ & $0.1\%$ &$0.81$\\
    \hline
    Overall & $26.6\%$ & $70.3\%$ & $2.9\%$ & $0.1\%$ & $0.0\%$ &$0.77$\\
    \hline
  \end{tabular}
\end{table}

\begin{table}
  \caption{\label{tab:NHSLS-mixing-sex} Reported mixing matrix, by sex}
  \centering
  \begin{tabular}{llrrr}
    \hline
    & & \multicolumn{2}{c}{Alter} & \\
    \cline{3-4}
    & & Female & Male & Total \\
    \hline
    \multirow{2}{*}{Ego} & Female & $0.4\% $ & $53.0\%$ & $53.5\%$  \\
    & Male & $45.5\%$ & $1.0\%$ & $46.5\%$  \\
    \hline
    &Total & $45.9\%$ & $54.1\%$ & $100.0\%$ \\
    \hline
  \end{tabular}
\end{table}

\subsubsection{Race}
We model race effects as an overall propensity of each racial category
to have partners and as a propensity to have partners in the same
racial category (\citealp{mcpherson2001bfh}; also see Table~\ref{tab:NHSLS-mixing-race}), with the following ERGM
statistics:
\begin{description}
\item[overall propensity of actors of each race to have
  ties] represented by the number of partners of actors in each racial
  category but one:
  \[g_{[6,7,8]}(y,\attrY)=\left(\sum_{i\in\text{Hispanic}}\abs{ y_{i}},
  \sum_{i\in\text{Other}}\abs{ y_{i}},\sum_{i\in\text{White}}\abs{
    y_{i}}\right),\] with category \enquote{Black} used as an arbitrary
  baseline for alphabetical reasons;
\item[race homophily] represented by the number of ties within each racial category: 
\begin{multline*}g_{[9,10,11,12]}(y,\attrY)=\left(\abs{ y_{\text{Black},\text{Black}}},\abs{
  y_{\text{Hispanic},\text{Hispanic}}},\right.\\ \left.\abs{
  y_{\text{Other},\text{Other}}}, \abs{
  y_{\text{White},\text{White}}}\right).
\end{multline*}
\end{description}

\begin{table}  
  \caption{\label{tab:NHSLS-mixing-race} Reported mixing matrix, by racial category}
  \centering
    \begin{tabular}{llrrrrr}
      \hline
    & & \multicolumn{4}{c}{Alter} & \\
    \cline{3-6}
    & & Black & Hispanic & Other & White & Total \\
    \hline
    \multirow{4}{*}{Ego} 
    & Black & $15.2\% $ & $0.3\%$ & $0.2\%$ & $0.5\%$ & $16.2\%$ \\
    & Hispanic & $0.2\% $ & $6.0\%$ & $0.3\%$ & $3.2\%$ & $9.7\%$ \\
    & Other & $0.0\% $ & $0.1\%$ & $2.5\%$ & $0.7\%$ & $3.3\%$ \\
    & White & $0.8\% $ & $1.5\%$ & $1.0\%$ & $67.6\%$ & $70.8\%$ \\
    \hline
    &Total & $16.2\% $ & $7.8\%$ & $4.0\%$ & $72.0\%$ & $100.0\%$ \\
    \hline
  \end{tabular}
\end{table}

\subsubsection{Age}
\begin{figure}
\noindent
\begin{center}
  \includegraphics[width=.49\textwidth,keepaspectratio]{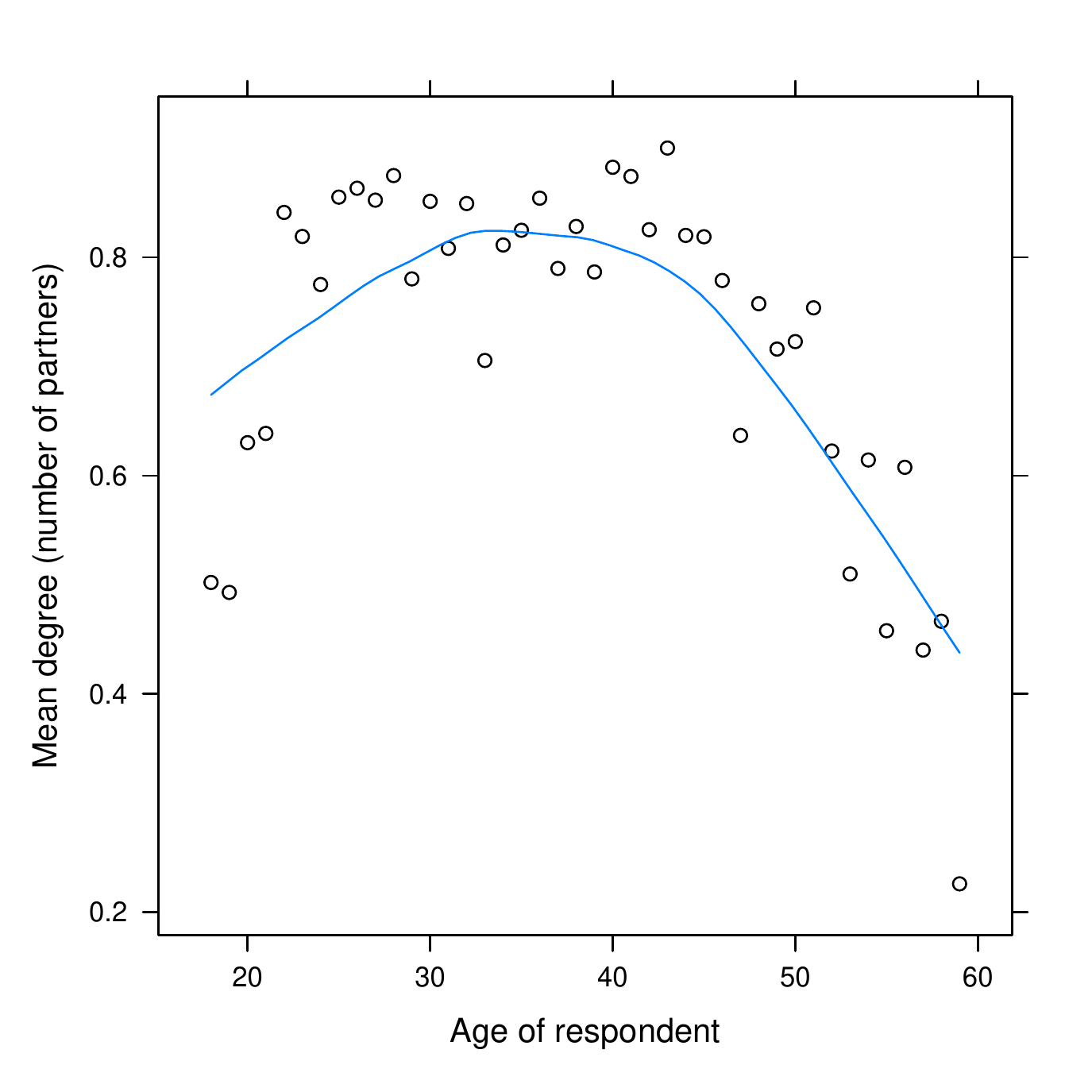}\includegraphics[width=.49\textwidth,keepaspectratio]{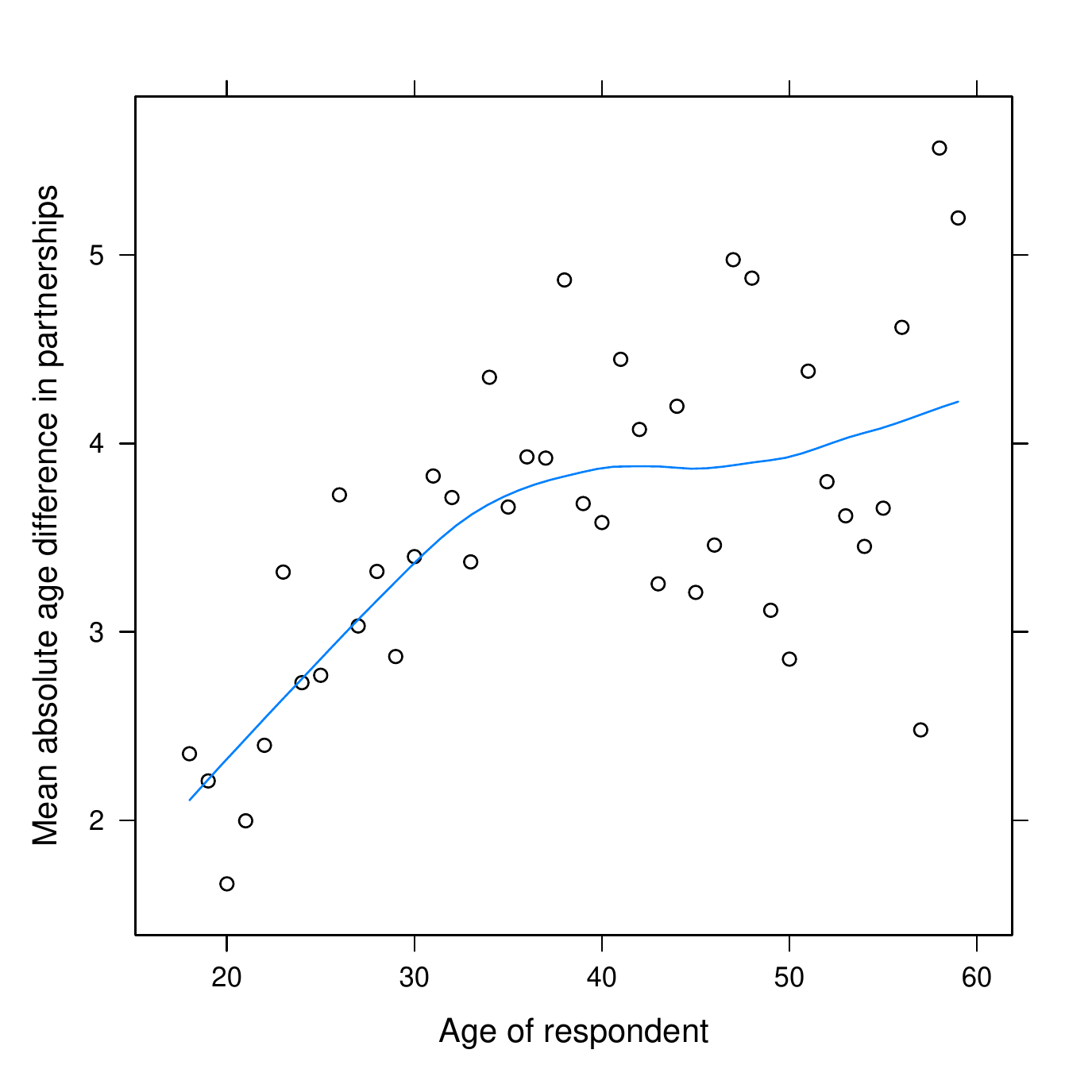}
\end{center}
\caption[Mean degree and age difference as a function of
age]{\label{fig:NHSLS-age-effects} Mean number of partners reported by a
  respondent (left) and mean absolute difference between age of
  respondent and that of the respondent's partner(s) (right), as a
  function of age}
\end{figure}

We model the effect of age on tie probabilities in three ways. (We
illustrate them in Figure~\ref{fig:NHSLS-age-effects}.) Firstly,
actors at different ages may have different overall propensities for
partnerships, and, furthermore, the effect of the age on the number of
partners would be stronger for younger actors. We thus model the
marginal effect of age as a quadratic function of the square root of
age. Secondly, actors tend to have partners of similar age
\citep{mcpherson2001bfh}, again, with the effect being stronger for
younger ages. We thus model this effect with a quadratic function of
the difference between ages of partners and a quadratic function of
the difference between square roots of their ages. Lastly, there is
age asymmetry in heterosexual relationships, described above.

To reduce correlations and improve the numeric conditioning of the
model, we center and scale the ages of actors to be between
$-\frac{1}{2}$ and $+\frac{1}{2}$. The transformation used does not
modify the model itself, only the coefficients. This results in the
following ERGM statistics:
\begin{description}
\item[overall age effects] represented by the summing, over all the
  actors, the product of each actor's number of partners and of each
  function of interest of that actor's age:
 \[g_{[13,14]}(y,\attrY)=\left(\sum_{i=1}^n \abs{ y_i} \left(\sqrt{\frac{t_i-18}{60-18}}-\frac{1}{2}\right),\sum_{i=1}^n \abs{ y_i} \left(\frac{t_i-18}{60-18}-\frac{1}{2}\right)\right);\] 
\item[age difference effects] represented by the summing, over all the
  dyads, the product of the value of each dyad and of each function
  of interest of the incident actors' ages:
  \begin{multline*}g_{[15,16,17,18]}(y,\attrY)=\left(\sum_{\pij\in \dysY} \yij \abs{\sqrt{\frac{t_i-18}{60-18}}-\sqrt{\frac{t_j-18}{60-18}}}^p,\vphantom{\sum_{\pij\in \dysY} \yij \abs{ \frac{t_i-18}{60-18} - \frac{t_j-18}{60-18}}^p}\right.\\
    \left.\vphantom{\sum_{\pij\in \dysY} \yij \abs{\sqrt{\frac{t_i-18}{60-18}}-\sqrt{\frac{t_j-18}{60-18}}}^p}\sum_{\pij\in \dysY} \yij \abs{ \frac{t_i-18}{60-18} - \frac{t_j-18}{60-18}}^p\right)_{p\in\{1,2\}}.\end{multline*}
\end{description}

\subsection{Simulation study design}
We performed two simulation studies. Firstly, we compared parameter
estimates for 400 egocentric bootstrapped resample sizes ranging from
600 to 12,000, logarithmically spaced. Secondly, we compared 100
resamples of each of the sizes 1,000, 6,000, and 11,000. For each
sample size, we generated estimates as follows:
\begin{itemize}
  \item[1)] Resample the desired numbers egos and their alters from
    the NHSLS dataset, as described in Section~\ref{NHSLS-weighting}.
  \item[2)] Compute network statistics as described in Section~\ref{sec:ego-meanstats}.
  \item[3)] Fit an ERGM with terms and offset described in Section~\ref{sec:NHSLS-model} using \proglang{R} \citep{proglangrdevelopmentcoreteam2009rle} package
    \pkg{statnet} \citep{handcock2008sst}.
  \item[4)] Record the ERGM parameter estimates $\hat{\theta}$.
\end{itemize}
In the framework of \citet*{duijn2009fcm}, we are considering networks
of a given size with mean value parameters derived as described in
Section~\ref{sec:egodata}, and consider whether the corresponding
natural parameters, in the presence of an offset, are invariant to
network size. A model with good invariance properties would thus
produce natural parameter estimates that do not change substantially
with a changing network size. The variability in the estimates due to
bootstrap resampling provide a baseline for the magnitude of this
change.

\subsection{\label{sec:NHSLS-results}Results}
We give the estimated model coefficients for the three-resample-size
simulation in Table~\ref{tab:NHSLS-results}. The model parameter
estimates are consistent with our expectations: same-sex partnership
count has a significant negative coefficient, and there is a strong
bias for monogamy for both sexes. Race homophily is consistently
positive. The positive coefficient on the age asymmetry term indicates
a bias for older-male-younger-female partnerships as well.

\begin{table}
  \caption{\label{tab:NHSLS-results} Average bootstrap estimates (and
    standard errors) of NHSLS model parameters, by sample size}
  \centering
  \begin{tabular}{lrrr}
    \hline
  & $N=1000$ & $N=6000$ & $N=11000$ \\
  Term & &&\\
  \hline
  Offset & $-6.91$ (fixed) & $-8.70$ (fixed) & $-9.31$ (fixed) \\
  \hline
  Actor activity by sex &&& \\
  \quad Female & $-1.29$ $(0.88)$ & $-1.10$ $(0.17)$ & $-1.19$ $(0.13)$ \\
  \quad Male & $-0.43$ $(0.91)$ & $-0.58$ $(0.16)$ & $-0.63$ $(0.12)$ \\
  Same-sex partnership & $-4.59$ $(1.75)$ & $-4.07$ $(0.14)$ & $-4.09$ $(0.11)$ \\
  Monogamy by sex &&&\\
  \quad Female  & $2.31$ $(0.33)$ & $2.17$ $(0.11)$ & $2.20$ $(0.08)$ \\
  \quad Male & $2.00$ $(0.25)$ & $1.93$ $(0.07)$ & $1.94$ $(0.05)$ \\
  Actor activity by race &&&\\
  \quad Black &$ 0 $ (baseline) &$ 0 $ (baseline)&$ 0 $ (baseline)\\
  \quad Hispanic & $0.86$ $(0.35)$ & $1.00$ $(0.13)$ & $1.02$ $(0.11)$ \\
  \quad Other & $1.28$ $(0.44)$ & $1.39$ $(0.16)$ & $1.42$ $(0.14)$ \\
  \quad White & $0.51$ $(0.45)$ & $0.58$ $(0.22)$ & $0.60$ $(0.15)$ \\
  Race homophily by race &&& \\
  \quad Black & $4.65$ $(0.60)$ & $4.85$ $(0.27)$ & $4.83$ $(0.18)$ \\
  \quad Hispanic & $2.84$ $(0.48)$ & $2.67$ $(0.22)$ & $2.70$ $(0.16)$ \\
  \quad Other & $3.34$ $(0.54)$ & $3.20$ $(0.21)$ & $3.17$ $(0.16)$ \\
  \quad White & $2.11$ $(0.42)$ & $2.09$ $(0.20)$ & $2.13$ $(0.15)$ \\
  Age effects &&&\\
  \quad $\sqrt{\text{age}}$ effect  & $-1.71$ $(0.54)$ & $-1.80$ $(0.24)$ & $-1.74$ $(0.18)$ \\
  \quad $\text{age}$ effect & $1.62$ $(0.44)$ & $1.65$ $(0.18)$ & $1.60$ $(0.13)$ \\
  Age difference effects &&&\\
  \quad Diff. in $\sqrt{\text{age}}$ & $-8.29$ $(2.38)$ & $-8.31$ $(1.21)$ & $-8.19$ $(0.96)$ \\
  \quad Diff. in $\text{age}$ & $-7.70$ $(2.24)$ & $-6.72$ $(1.05)$ & $-6.61$ $(0.74)$ \\
  \quad Squared diff. in $\sqrt{\text{age}}$  & $4.37$ $(3.27)$ & $4.08$ $(2.07)$ & $3.72$ $(1.80)$ \\
  \quad Squared diff. in $\text{age}$ & $6.05$ $(2.98)$ & $4.05$ $(1.51)$ & $3.77$ $(1.11)$ \\
  \quad Age-sex asymmetry & $0.98$ $(0.10)$ & $0.94$ $(0.04)$ & $0.95$ $(0.03)$ \\
\hline
\end{tabular}
\end{table}

More importantly, the parameter estimates are essentially stable as
the resampled network size ranges from 6,000 to 11,000. In fact, on
average, the difference between a parameter's estimate for 6,000 and
11,000 is smaller than 1 simulated standard error for that estimate
based on resample size of 11,000. These standard errors are \emph{not}
conservative, since the original dataset is one third of that size and
the resampling is reweighted, both of which lead to the standard
errors in Table~\ref{tab:NHSLS-results} being smaller than what the
standard deviations of the parameter estimates would be in a
hypothetical simple random sample of 11,000 egos. In short, the
difference in the estimates due to the difference in network size is
smaller than the differences due to sampling error.

\begin{figure}
\noindent
\begin{center}
\includegraphics[width=.99\textwidth,keepaspectratio]{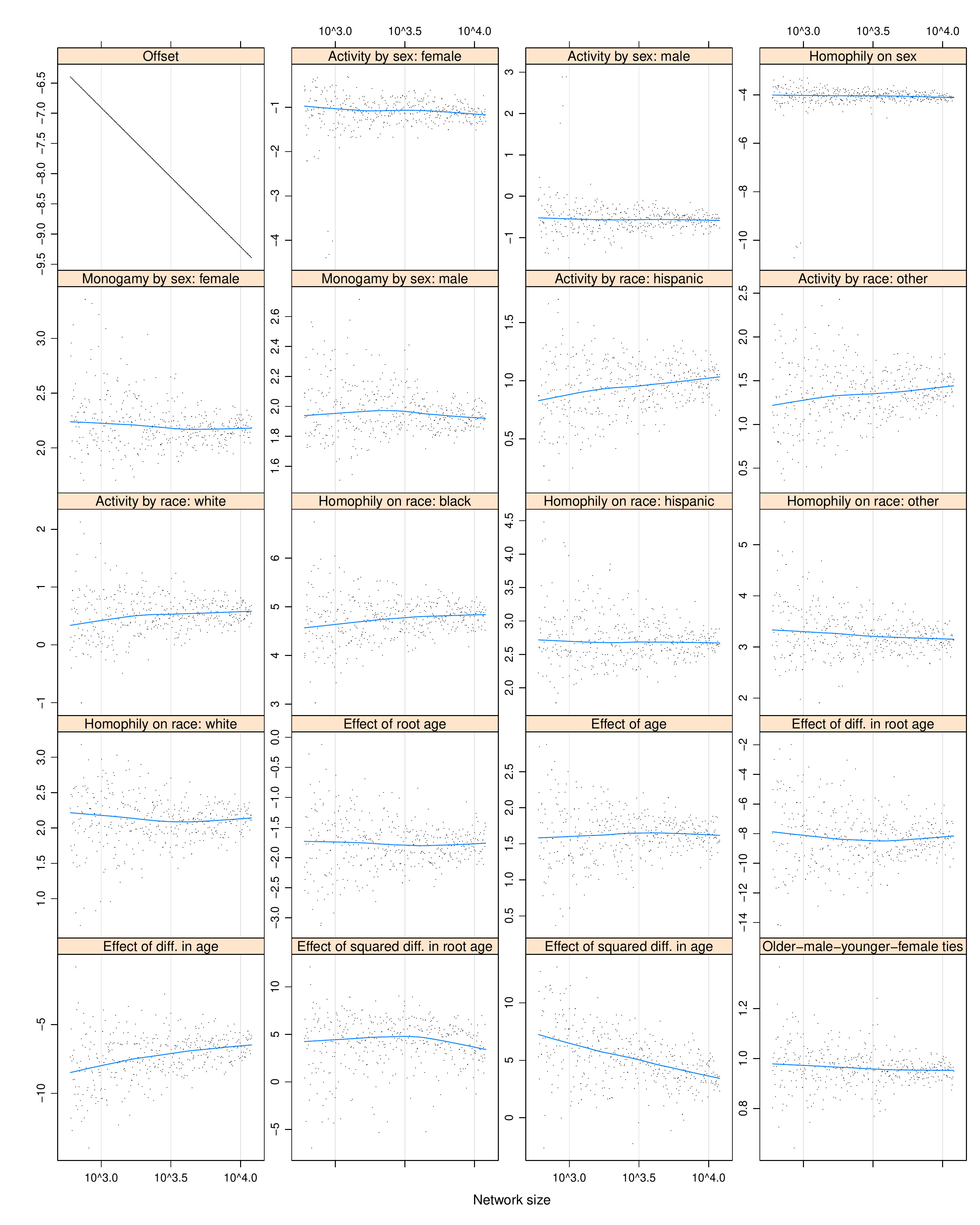}
\end{center}
\caption[Parameter estimates of the model fit to the resamples of
  NHSLS dataset, as a function of network
  size]{\label{fig:NHSLS-plot-coef} Parameter estimates of the
  model fit to the resamples of NHSLS dataset, as a function of
  network size, varying from 600 to 12,000, spaced
  logarithmically. Note that the horizontal axis is on a logarithmic
  scale.}
\end{figure}

The simulation of network sizes 600 through 12,000 shows a similar
trend. We give the trends in the parameter estimates in
Figure~\ref{fig:NHSLS-plot-coef}. The estimates show a pattern of
asymptoting, although asymptoting for some of them --- particularly
age difference effects --- appears to be slower. This could be an
artifact of normalizing the ages.

All this suggests asymptotic invariance to network size for this,
fairly complex, model. While limitations of computing capacity
preclude computing parameter estimates for network sizes in the
hundreds of millions --- the population of the United States in 1992 in
the age range surveyed --- it is likely that these estimates would be
very close to those for network size 12,000.

\section{\label{sec:discussion}Discussion}
Effect of network size and composition on ERGMs has received limited
attention in the literature to date. We have described the desired
behavior we would like a social network model to exhibit when applied
to social networks of different sizes, and have shown which of these
are (or are not) properties of an unadjusted ERGM. We propose a simple
adjustment based on an offset term that appears to produce,
asymptotically but also for networks of moderate size, the desired
behavior: given that the network statistics are \enquote{local} in
nature, the model produces the size appropriate mean statistics for
group mixing and degree distributions under a variety of network sizes
--- similar distributions map to similar parameter values.

We demonstrated this property by fitting a fairly complex ERGM to networks of different sizes constructed to have similar structure.

We also described an approach to fitting ERGMs to egocentrically
sampled data that makes use of our heuristics for similar structure
across network sizes. Combined with the proposed network size
adjustment for ERGMs, it may allow the parameter estimates from
fitting the network data on a sample to be generalized to the
population from which the sample was drawn.

This approach provides a principled framework for network comparison
and simulation. With the offset adjustment, the remaining parameters
in the ERGM can be used to test whether network structures represented
in the model are statistically different between two networks, even if
the networks have different size and/or composition.  Since the
parametrization is now size and composition invariant, this approach
can also be used to simulate networks that have the same underlying
structure, though they may have different size and composition.

Our analysis leaves open some questions. We limit our statistics to
first- and second-order effects --- dyadic and degree distribution
effects --- and do not discuss third-order effects such as
triad-closure bias, and while two of the notions of locality that we
describe allow modeling of such effects, we do not examine the
properties of these in the presence of an offset, because the
egocentrically sampled data available do not contain information about
such effects. \citet{goodreau2008bff}, using dyad census data,
 fit a geometrically-weighted edgewise shared
partner (GWESP) statistic
\citep{snijders2006nse,hunter2006ice,hunter2007cef}, used to model
triad-closure bias, and found that the coefficient of the GWESP
statistic appeared to asymptote as school sizes increased. Other
parameters, except for the overall density parameter, also did not
appear to depend on school sizes \citep{goodreau2009rap}. This suggests
that our approach applies to third-order effects as well, but this is
a subject for future research.

Another question is whether convergence to the asymptotic estimates
could be sped up by modifying the offset term: $\abs{y}\log
\frac{1}{n}$ has the advantage of simplicity, but there are other
candidates, such as $\abs{y}\logit\left(\frac{\mu}{n-1}\right)$, for a
constant $\mu\ll n$ that may have better properties. At the same time,
if the sufficient statistics of the model or some linear combination
thereof include $\abs{y}$, the change in the offset coefficient will
be absorbed only into their parameter estimates, and will not affect
the convergence of the others. In our example, the number of partners
of males and the number of partners of females sum to twice the number
of edges, and, thus, their coefficients would play this role.

While we describe a way to use egocentrically sampled data to
construct networks with similar structures but varying sizes, and
describe how these parameter estimates may be generalized to the
underlying network, we do not have an appropriate measure of
uncertainty of these estimates --- we note above that the standard
errors we report for the larger network sizes are too small ---
rigorously assessing this uncertainty is a subject of ongoing work.

Lastly, network size can affect the structure of a network in ways
other than density, and we do not explore these effects here.

\section*{Acknowledgments}
\addcontentsline{toc}{section}{Acknowledgments}

The authors wish to acknowledge the following grants as having
supported this research: NSF Grant SES-0729438 and NIH Grants HD-41877
and DA-12831 (all authors); NSF Grant MMS-0851555 and ONR Award
N00014-08-1-1015 (Mark Handcock); and Portuguese Foundation for
Science and Technology Ci\^{e}ncia 2009 Program and a grant from the
NSA to the Department of Statistics at the University of Washington
(Pavel Krivitsky).

\bibliography{Adjusting_for_Network_Size_and_Composition_Effects_in_ERGMs}
\addcontentsline{toc}{section}{References}
\bibliographystyle{plainnat}

\appendix
\section*{Appendix}
\section{\label{app:change-stats}Change statistics and Gibbs sampling}
\subsection[Conditional probability of a dyad (i,j)]{\label{app:change-stat-deriv}Conditional probability of a dyad $\pij$}
ERGM distribution in~\eqref{eq:cergm} has the conditional probability
of an edge at $\pij$, given the rest of the network, of
\begin{align*}
\Peg(&\Yij=1|\attrY,Y-\pij =y-\pij ;\theta)=\\
&=\frac{\Peg(Y=y+\pij |N;\theta)}{\Peg(Y=y-\pij |N;\theta)+\Peg(Y=y+\pij|N;\theta)}\\
 &=\frac{\frac{\myexp{\innerprod{\eta(\theta,\attrY)}{g(y+\pij ,\attrY)}}}{\cancel{\ceg(\theta,\attrY)}}}{\frac{\myexp{\innerprod{\eta(\theta,\attrY)}{g(y-\pij ,\attrY)}}}{\cancel{\ceg(\theta,\attrY)}}+\frac{\myexp{\innerprod{\eta(\theta,\attrY)}{g(y+\pij ,\attrY)}}}{\cancel{\ceg(\theta,\attrY)}}}\\
  &=\frac{1}{\myexp{\innerprod{\eta(\theta,\attrY)}{g(y-\pij ,\attrY)}-\innerprod{\eta(\theta,\attrY)}{g(y+\pij ,\attrY)}}+1}\\
  &=\frac{1}{1+\myexp{-\innerprod{\eta(\theta,\attrY)}{\Delta\sij g(y,\attrY)}}}\\
  &=\ilogit\left(\innerprod{\eta(\theta,\attrY)}{\Delta\sij g(y,\attrY)}\right),
\end{align*}
where $\ilogit(x)\equiv \frac{1}{1+\myexp{-x}}$ and $\Delta\sij g_k(y,\attrY)\equiv g_k(y+\pij ,\attrY)-g_k(y-\pij ,\attrY)$.

\subsection{\label{app:Gibbs-alg}Na\"{i}ve Gibbs sampling algorithm for ERGMs}
The following algorithm can be used to generate a random draw from an ERGM probability distribution~\eqref{eq:cergm}
with an intractable normalizing constant:
\begin{algorithmic}[1]
  \REQUIRE Arbitrary $y^{0} \in \netsY$ and $S$ sufficiently large
  \FOR{$s\gets 1$ to $S$}
  \STATE $\pij \gets \text{\textsc{RandomChoose}}(\dysY)$
  \STATE\label{alg:Gibbs-alg-r} $r \gets \ilogit\left(\innerprod{\eta(\theta,\attrY)}{\Delta\sij g(y,\attrY)}\right)$ \COMMENT{i.e. $\Peg(\Yij=1|\attrY,Y^{(s-1)}-\pij=y^{(s-1)}-\pij;\theta)$}
  \STATE $u \gets \Uniform(0,1)$
  \IF{$ u<r $}
  \STATE $y^{s} \gets y^{(s-1)}+\pij$ \COMMENT{Have a tie at $\pij$ with probability $r$.}
  \ELSE
  \STATE $ y^{s} \gets y^{(s-1)}-\pij$ \COMMENT{Have no tie at $\pij$ with probability $1-r$.}
  \ENDIF
  \ENDFOR
  \RETURN $y^{S}$
\end{algorithmic}

Here, $\text{\textsc{RandomChoose}}(A)$ is a function that, given a
set, $A$, selects and returns a member $a\in A$ at random.

\section{Details of asymptotic properties of ERGMs}
\subsection{\label{app:scale-ergm-proof}Size-invariant statistics of linear ERGMs}
In this section, we prove the assertion about linear ERGMs that was
stated in Section~\ref{sec:scale-ergm}. Consider a sequence of random
undirected networks $Y\ofsize{n_1},Y\ofsize{n_2},\dotsc$ of increasing size
whose actor attributes $\attrY\ofsizen$ such that frequency or
distribution of any exogenous actor attributes converges as
$n\to\infty$ --- that is, the network size grows, but the composition
does not change. Because we do not model actor attributes in this
discussion, an intuitive way to construct such a sequence is by
defining some initial set of actor attributes $\attrY\ofsize{n_0}$ and
defining, for any integer $k>1$, $\attrY\ofsize{n_k}$ to simply be
$\attrY\ofsize{n_0}$ replicated $k$ times.

Let $\eta(\theta,\attrY\ofsizen)=\theta$ be the natural parameter
vector (i.e. a linear ERGM); $g(\cdot,\cdot)$ be a vector of network
statistics that \emph{may} also depend on network size and
composition; and $T(\cdot,\cdot)$ be the vector of network statistics,
which may depend on network size and composition, but whose expected
value needs to remain constant as network size changes or converge to
a finite limit as it increases.

Suppose that for some $T(\cdot,\cdot)$ of interest,
\begin{multline}
  \exists_{g(\cdot,\cdot)}\forall_{\theta}\forall_{\epsilon>0}\exists_{N<\infty}\forall_{n>N,n'>N}\left\lvert\sum_{y\in\netsY\ofsizen}T(y,\attrY\ofsizen)\frac{\myexp{\theta \cdot g(y,\attrY\ofsizen)}}{\ceg(\theta,\attrY\ofsizen)}-\vphantom{\sum_{y\in\netsY\ofsize{n'}}T(y,\attrY\ofsize{n'})\frac{\myexp{\theta \cdot g(y,\attrY\ofsize{n'})}}{\ceg(\theta,\attrY\ofsize{n'})}}\right.\\
  \left.\vphantom{\sum_{y\in\netsY\ofsizen}T(y,\attrY\ofsizen)\frac{\myexp{\theta \cdot g(y,\attrY\ofsizen)}}{\ceg(\theta,\attrY\ofsizen)}-}\sum_{y\in\netsY\ofsize{n'}}T(y,\attrY\ofsize{n'})\frac{\myexp{\theta \cdot g(y,\attrY\ofsize{n'})}}{\ceg(\theta,\attrY\ofsize{n'})}\right\rvert<\epsilon.\label{eq:size-no-effect}
\end{multline}
That is, suppose that for this particular $T(\cdot,\cdot)$, there
exists a vector of statistics $g(\cdot,\cdot)$ such that for any given
fixed value of natural parameter vector $\theta$, the maximum
difference in expected value of the statistic of interest
$T(\cdot,\cdot)$ due to differences in network size can be made
arbitrarily small for sufficiently large networks. Then
\citep[p. 71]{strichartz2000wa}
\begin{equation}
  \exists_{g(\cdot,\cdot)}\forall_{\theta}\lim_{n\to\infty}\sum_{y\in\netsY\ofsizen}T(y,\attrY\ofsizen)\frac{\myexp{\theta \cdot g(y,\attrY\ofsizen)}}{\ceg(\theta,\attrY\ofsizen)}=t_g(\theta,\AttrY)<\infty:\label{eq:size-no-effect-lim}
\end{equation}
the expected value converges to some $t_g(\theta,X)$, a function of
$\theta$ and asymptotic network composition distribution $\AttrY$.
For this particular combination of $g(\cdot,\cdot)$ and
$T(\cdot,\cdot)$,~\eqref{eq:size-no-effect-lim} holds for all
$\theta$, and therefore holds for $\theta=0$. But then,
\begin{align*}
  &\lim_{n\to\infty}\sum_{y\in\netsY\ofsizen}T(y,\attrY\ofsizen)\frac{\myexp{0 \cdot g(y,\attrY\ofsizen)}}{\ceg(0,\attrY\ofsizen)}=t_g(0,\AttrY),\\
  &\lim_{n\to\infty}\sum_{y\in\netsY\ofsizen}T(y,\attrY\ofsizen)\frac{1}{\sum_{y'\in\netsY_{n}}1}=t_g(0,\AttrY),\\
  &\lim_{n\to\infty}\frac{1}{\abs{\netsY\ofsizen}}\sum_{y\in\netsY\ofsizen}T(y,\attrY\ofsizen)=t_g(0,\AttrY).
\end{align*}
This summation is just the expected value of $T(\cdot,\cdot)$
under an \ErRe{} graph of that size and composition with each dyad
value having an independent $\Bernoulli\left(\half\right)$
distribution, with $\attrY\ofsizen$ being irrelevant, so
\[\lim_{n\to\infty}\E\left(T(Y,\attrY\ofsizen)\right)=t_g(0,\AttrY)\]
where $Y$ is a $\Bernoulli\left(\half\right)$ graph of size $n$.

Thus, regardless of what $g(\cdot,\cdot)$ may be, unless
$T(y,\attrY\ofsizen)$ already has the property of its expectation
converging as its network size increases (at least in a Bernoulli
model), it is not possible to construct an ERGM that
satisfies~\eqref{eq:size-no-effect}. In particular, the expected mean
degree in an undirected $\Bernoulli(\half)$ graph of size $n$ is
$\frac{n-1}{2}\to \infty$ as $n\to \infty$, so the degree distribution
does not remain unaffected or converge under changing network
size. The network statistics that \emph{can} be made unaffected by
network size include the density of the network, the densities of
subnetworks, and affine transformations thereof with fixed
coefficients.

\subsection{\label{app:erre-asymp}Asymptotic degree distribution of a simple offset model}

Starting with \eqref{eq:erre-offset}, let
\[p_n=\Peg(\Yij\ofsizen=1|\attrY\ofsizen;\theta)=\ilogit\left(-\log n + \theta\right).\]
Then,
\begin{align*}
\lim_{n\to\infty}&\Peg(|Y\ofsizen_{i}|=d|\attrY\ofsizen)\\
&=\lim_{n\to\infty}\binom{n-1}{d}\left({p_n}\right)^d\left(1-{p_n}\right)^{n-1-d}\\
&=\lim_{n\to\infty}\frac{(n-1)!}{(n-1-d)!d!}\left(\myexp{\logit{p_n}}\right)^d\left(1-{p_n}\right)^{n-1}\\
&=\lim_{n\to\infty}\frac{\prod_{k=1}^d (n-k)}{d!}\left(\myexp{-\log n + \theta}\right)^d\\
&\phantom{=\lim_{n\to\infty}}\times\left(1-\frac{1}{1+n\myexp{-\theta}}\right)^n \left(1-\frac{1}{1+n\myexp{-\theta}}\right)^{-1}\\
&=\lim_{n\to\infty}\left(\prod_{k=1}^d \frac{n-k}{n}\right)\frac{1}{d!}\left(\myexp{\theta}\right)^d\\
&\phantom{=\lim_{n\to\infty}}\times\left(1+\frac{\myexp{\theta}}{n}\right)^{-n} \left(1-\frac{1}{1+n\myexp{-\theta}}\right)^{-1}\\
&=\frac{1}{d!}\left(\myexp{\theta}\right)^{d}\myexp{-\myexp{\theta}},
\end{align*}
the PDF of a Poisson distribution with mean $\myexp{\theta}$.
\end{document}